\documentclass[letterpaper, 10 pt, conference]{ieeeconf}  

\IEEEoverridecommandlockouts                              

\usepackage{graphics} 
\usepackage{epsfig} 
\usepackage{mathptmx} 
\usepackage{times} 
\usepackage{amsmath} 
\usepackage{amssymb}  

\usepackage{todonotes}
\usepackage{xcolor}
\usepackage{xspace}
\usepackage{svg}
\usepackage{booktabs} 
\usepackage{cite}
\usepackage{hyperref}

\title{\LARGE \bf
Crowdsourcing eHMI Designs: A Participatory Approach to Autonomous Vehicle-Pedestrian Communication
}

\author{Ronald Cumbal$^{1}$ and Didem G{\"u}rd{\"u}r Broo$^{1}$ and Ginevra Castellano$^{1}$
\thanks{$^{1}$Department of Information Technology,
        Uppsala University, Sweden. Mail correspondence to: 
        {\tt\small ronald.cumbal@it.uu.se}}%
}

\begin{document}

\maketitle
\thispagestyle{plain}
\pagestyle{plain}

\begin{abstract}
As autonomous vehicles become more integrated into shared human environments, effective communication with road users is essential for ensuring safety. While previous research has focused on developing external Human-Machine Interfaces (eHMIs) to facilitate these interactions, we argue that involving users in the early creative stages can help address key challenges in the development of this technology. To explore this, our study adopts a participatory, crowd-sourced approach to gather user-generated ideas for eHMI designs. Participants were first introduced to fundamental eHMI concepts, equipping them to sketch their own design ideas in response to scenarios with varying levels of perceived risk. An initial pre-study with 29 participants showed that while they actively engaged in the process, there was a need to refine task objectives and encourage deeper reflection. To address these challenges, a follow-up study with 50 participants was conducted. The results revealed a strong preference for autonomous vehicles to communicate their awareness and intentions using lights (LEDs and projections), symbols, and text. Participants' sketches prioritized multi-modal communication, directionality, and adaptability to enhance clarity, consistently integrating familiar vehicle elements to improve intuitiveness.
\end{abstract}

\section{Introduction}

Effective communication between Autonomous Vehicles (AVs) and road users is crucial for safe interactions in shared spaces \cite{merat2018externally}. Research indicates that pedestrians often expect feedback or acknowledgment from AVs, much like when interacting with human drivers \cite{schieben2019designing}. To meet this need, researchers have focused on developing external Human-Machine Interfaces (eHMIs) to facilitate communication between AVs and pedestrians.
While some studies have adopted participatory approaches ---particularly involving vulnerable populations \cite{colley2019including, colley2020towards, asha2021co}---most eHMI designs have been created primarily by experts. However, we believe involving end users earlier in the design process can offer valuable insights. 
By drawing on their collective real-world experiences, users can help address research gaps, such as communication ambiguities and the need for universal eHMI designs \cite{dey2020taming}.

To address these gaps, our study introduces a web-based participatory crowdsourcing method that bridges the scale advantages of crowdsourcing with the depth of participatory design. Unlike surveys that evaluate existing designs or small-scale workshops with fewer participants, our approach enables both quantitative preference data collection and qualitative creative input from a diverse user pool. 
Our goal is to develop an approach that not only captures users’ \textit{creative} input but also provides \textit{insights} into the thought processes and needs that shape their design choices.

\begin{figure}[t]
  \centering
  \includegraphics[width=0.49\textwidth]{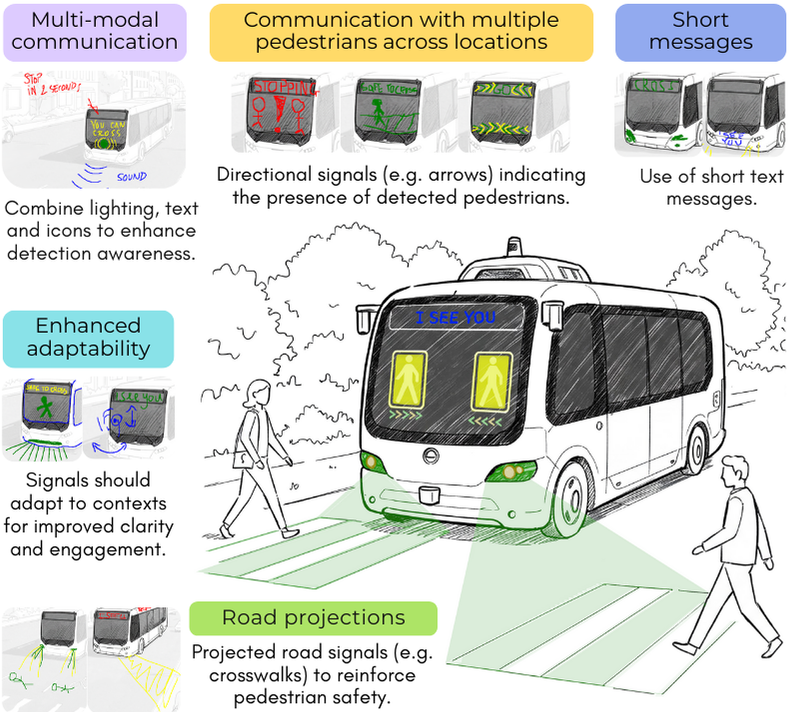}
  \caption{Illustration of the final design components proposed by participants.}
  \label{fig:teaser}
\end{figure}

This approach offers several distinct advantages over existing methods. It can scale to larger participant pools than traditional co-design workshops, provides richer creative input than surveys, and captures the reasoning behind design preferences rather than just the preferences themselves. We analyze user responses across different risk levels, comparing low-risk scenarios (marked pedestrian crossings with signals) to high-risk situations (unmarked crossings with multiple pedestrians). This systematic variation of risk contexts reveals how everyday users adapt their communication expectations based on situational factors, informing context-aware design principles for future eHMI development.

While surveys and cross-cultural evaluations have been used in prior research, these methods often focus primarily on evaluation \cite{verma2019pedestrians}. 
In contrast, our approach integrates user insights from the outset, actively involving them in shaping eHMI design. We acknowledge that participatory methods are beneficial for engaging vulnerable populations, though our approach does not specifically cater to these groups. Nevertheless, collecting input from a larger participant pool allows us to better understand user expectations while gaining meaningful insights into how everyday people engage with technology. 
Our main contribution is a scalable method that combines participatory approaches with crowdsourcing to explore how pedestrians envision eHMI designs, paving the way for more universal AV communication.

\section{Related Work}
The integration of eHMIs into autonomous vehicles has attracted substantial research interest in recent years. While some debate their necessity \cite{deWinter2022external}, the need for clear communication between vehicles and road users remains a stronger argument \cite{rasouli2019autonomous}. 
Research has explored various ways for autonomous vehicles to convey information, with one of the main challenges being the need for universal design \cite{dey2020taming}. 
Participatory methods can provide valuable insights for this challenge, as they are particularly effective in aligning technology with real-world user needs. 
However, despite their use in some studies, they remain underutilized, as expert-driven approaches continue to dominate eHMI development.

Several studies highlight the importance of participatory approaches in designing effective eHMIs. The EU Project CityMobil2, for instance, used interviews, onsite questionnaires, and focus groups to assess pedestrian and cyclist perceptions of automated minibuses \cite{schieben2019designing}, leading to specific design recommendations for vehicle-to-pedestrian communication. 
Similarly, Verma et al. \cite{verma2019pedestrians} examined visual communication strategies for autonomous shuttles, comparing instructional, anthropomorphic, metaphorical, and symbolic signals. Their findings, based on an initial participatory forum with twelve participants, showed that anthropomorphic signals were most effective in capturing attention and boosting pedestrian confidence.

Beyond general design insights, participatory approaches also help uncover perspectives from underrepresented groups, a recognized priority in eHMI research \cite{colley2019including}. For example, Asha et al. \cite{asha2021co} conducted a five-week remote co-design study with a researcher who uses a powered wheelchair, revealing challenges such as visual obstructions in traffic. Colley et al. \cite{colley2020towards} facilitated a workshop with six visually impaired participants, identifying the need for clearer messaging and multi-vehicle communication. A follow-up virtual reality study reinforced these findings, showing that explicit, high-content messages improved trust, reduced cognitive load, and enhanced perceived safety.

A closely related study by Mahadevan et al. \cite{mahadevan2018communicating} employed the PICTIVE participatory design method, where twelve participants created 34 unique interface concepts for conveying an AV's awareness and intent. The results emphasized LED lights and LCD displays while also considering auditory and physical cues for visually impaired pedestrians. Chauhan et al. \cite{chauhan2024transforming} explored an infrastructure-based eHMI concept, gathering insights from a think tank discussion with 25 participants about a smart pole interaction unit.

While these studies provide valuable insights, they typically involve small participant groups. In contrast, crowdsourcing methods can leverage large-scale online participation to expand the diversity and scope of user input \cite{geiger2014personalized}. Although previous research has explored crowdsourcing in design \cite{xu2015classroom, park2013crowd}, its use in emerging technologies and facilitating creative tasks remains relatively uncommon \cite{oppenlaender2020creativity}.

This section highlights the benefits of participatory approaches but also underscores two key limitations: (1) scale constraints, as most studies engage only a small number of participants in the creative phase, limiting broader statistical analysis, and (2) diversity limitations, as co-located workshops tend to draw from local populations with similar cultural backgrounds. To address these gaps, our web-based crowdsourcing method enables asynchronous participation from diverse geographic locations, integrates learning modules to educate participants before design tasks, and includes reflective follow-up questions that encourage users to consider edge cases and refine their initial ideas.

\section{Research Objectives}
The previous section highlighted the benefits of participatory methods in providing deeper individual insights, while crowdsourcing excels at efficiently gathering diverse, large-scale data. This study combines both approaches to advance human-centered technology, guided by the following research questions:

\begin{itemize}
    \item \textbf{RQ1}: What insights can be learned from everyday user-inspired designs to inform the development of eHMIs?
    \item \textbf{RQ2}: How do pedestrians envision communication mechanisms in low- and high-risk traffic scenarios?
\end{itemize}

\begin{figure*}[t]
  \centering
  \includegraphics[width=1\linewidth]{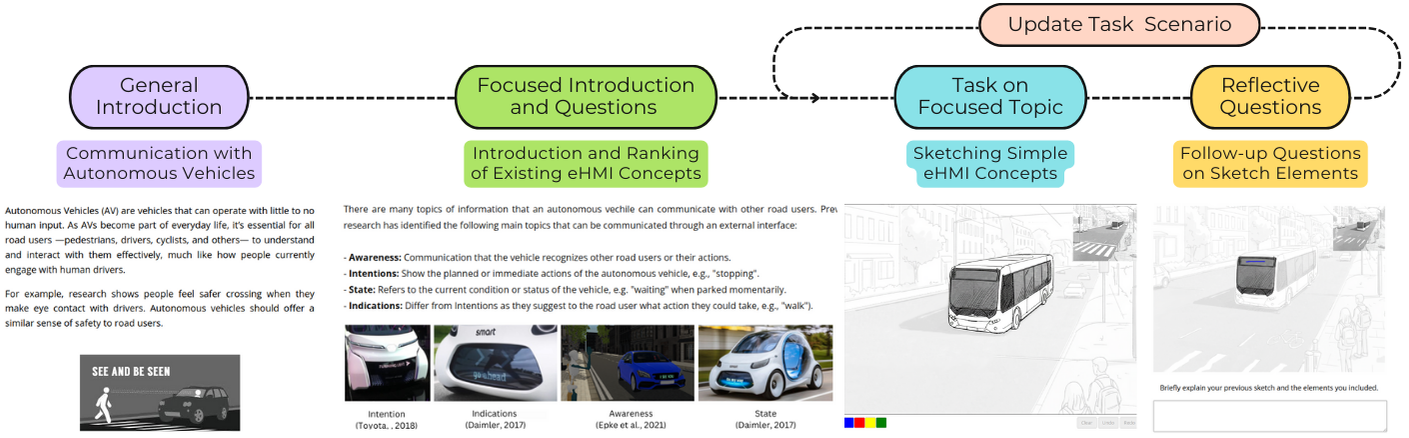}
  \caption{Diagram of the proposed method to crowdsource eHMI designs with screenshots of the implemented platform used to collect participant data.}
  \label{fig:diagram}
\end{figure*}

\section{Methodology}
The study began with a pilot test focused on two main aspects: (1) an interactive process designed to introduce participants to basic eHMI concepts and gather broad user preferences, and (2) a sketching task where participants created design ideas and explained their reasoning. 
A follow-up study was then developed to address identified limitations, with a particular focus on encouraging deeper reflection in participant responses.  
This section details the methodology used in the final study, with a summary of the key differences from the pilot test provided at the end. 
The experiment was implemented using the \textit{jsPsych} library, a tool designed for conducting behavioral experiments in web browsers \cite{deLeeuw2023jspsych}. 

\subsection{Introduction: Communication with Autonomous Vehicles}
At the beginning of the study, participants were provided with a brief overview of their task and the research objectives, following previous works \cite{dong2024exploring}. 
This was described as \textit{gathering insights from road users on how they envision communication with autonomous vehicles}.  
We also emphasized that \textit{their contributions could play a valuable role in shaping future guidelines and recommendations for eHMI design} to motivate participants.

Participants were then introduced to the definition of autonomous vehicles and the importance of natural communication between road users and human drivers, as well as their autonomous counterparts, as done in previous co-designing processes \cite{chauhan2024transforming, colley2019including}. 
To show the context of previous research, a collage of eHMI designs proposed in earlier studies was included. 
Following this introduction, participants completed a short questionnaire to provide demographic information, including their age, gender, and level of experience or familiarity with autonomous vehicles. 

\subsection{Ranking Survey: Introduction to Existing eHMI Designs}
Participants were then introduced to key concepts in eHMI design, including the types of information autonomous vehicles communicate (\textit{Awareness}, \textit{Intention}, \textit{Indications}, and \textit{State}) and the visual/auditory components used for this purpose. 
These components were consolidated from prior research \cite{bazilinskyy2019survey, rasouli2019autonomous} and included: \textit{Speech} (spoken messages), \textit{Audio} (sounds like bird chirps), \textit{LED Lights} (static or flashing), \textit{Projections} (road displays), \textit{Symbols} (icons or graphical representations), \textit{Text} (written phrases), and \textit{Human-like} features. 
Each component was explained in detail before participants ranked them by preference. 
To aid understanding, all ranking options were illustrated with images from a previously proposed eHMI design, as shown in Figure \ref{fig:diagram}. 

\subsection{Sketching: Designing eHMI Concepts}
In the last section, participants created freehand sketches of how they envisioned an autonomous vehicle communicating with them.
To familiarize them with the sketching interface, they were first shown a diagram explaining key features, such as the color palette and navigation buttons.
Next, they were introduced to two scenarios with increasing perceived risk levels\footnote{Risk was defined from the pedestrian's perspective rather than with an objective risk measurements, following prior eHMI research \cite{de2023predicting, ha2020effects}.}. These scenarios were designed using storyboards to enhance engagement in crowdsourcing tasks \cite{morschheuser2017gamified} and, described as follows:

\begin{itemize}
    \item \textbf{Low-Risk Scenario}: Imagine you are about to cross a street at a zebra crossing with a traffic light when an AV approaches the intersection. This is a low-risk situation, as AVs are generally trusted to follow traffic rules.\footnote{Previous studies have shown that people generally expect AVs to obey traffic signals at regulated intersections \cite{schieben2019designing}.}
    \item \textbf{High-Risk Scenario}: Now, imagine a riskier scenario. You are about to cross the street, but now there are more people, including children, but there are no clear traffic signals. This situation feels more uncertain. The autonomous vehicle is much closer, so you can only use the windshield to communicate your design.
\end{itemize}

As part of the methodology, participants were instructed to mentally place themselves within each scenario and reflect on the factors that would make them trust or hesitate before crossing (as done in \cite{schieben2019designing,dong2024exploring}). 
To stimulate creativity, they were explicitly encouraged to combine existing technologies or envision novel futuristic solutions. 
Each scenario was accompanied by a contextual image to aid visualization. 
For the sketching task, participants used an 896 × 640 pixel digital canvas with a four-color palette (blue, red, yellow, and green). 
The background image provided a zoomed-in view of the autonomous vehicle, which changed based on the scenarios: (1) in the first, it showed the entire vehicle, and (2) in the second, it focused on the windshield area (see Figure \ref{fig:diagram}). 
This narrowing of focus aligns with prior research showing that as a vehicle approaches a crossing where a pedestrian is standing, their gaze shifts from the vehicle’s surroundings to the expected location of the driver \cite{dey2019gaze}.

Participant sketches were stored as \textit{base64-encoded} strings to retain color and stroke data. During post-processing, they were decoded into PNG format for visual analysis and systematic coding of interface components.

\subsubsection{Follow-up Questions}
After completing each sketching task, participants were asked to briefly describe their designs while viewing their submitted sketches. 
This step helped clarify their ideas and facilitated the post-experiment analysis of each submission.  
For the low-risk scenario, these descriptions were especially important in tailoring follow-up questions that prompted participants to reflect on potential ambiguities or limitations in their designs. 
To achieve this, we processed the descriptions and matched keywords to relevant questions. 
For instance, if a participant mentioned audio-related words, they would receive a question about how these signals would remain effective in noisy environments. 
The keyword list was generated based on the most frequently and relevant words used in the pilot test. If no relevant keywords were detected, a default question was presented. 
For the high-risk scenario, participants received a fixed follow-up question asking how they would modify the system to communicate effectively with multiple pedestrians engaged in different activities ---a relevant and timely research topic \cite{chauhan2024transforming}. 
The complete list of follow-up questions is provided in Table \ref{tab:followupQuestions}. 
At the end of the study, participants had the opportunity to share any additional comments and were asked whether they would be interested in participating in similar tasks in the future.

\begin{table}[t]
\caption{Follow-up questions based on keyword-matching.}
\label{tab:followupQuestions}
\begin{tabular}{p{2cm}|p{5.8cm}} 
\textbf{Keywords} & \textbf{Question} \\ \hline
Projections, flash(ing), led(s), light(s), screen &
[a] How would you modify the visual displays to remain effective in different weather conditions (e.g., bright sunlight, fog, or nighttime) or in complex environments like busy urban streets? \\ \hline
Audio, voice(s), sound(s), loud, speaker, auditory &
[b] How would you adjust the sound signals to remain effective in noisy environments while avoiding noise that could disturb others?\\ \hline
Message(s), Text(s) &
[c] How would you adapt the text to ensure they are accessible to people who may have difficulty seeing, understanding, or reading them? \\ \hline
Eye(s), smile(s), see(s) &
[d] How would you add gestures (such as waving, blinking, or facial expressions) to make a clearer message? \\ \hline
[None] &
[f] How would you modify the system to communicate with multiple pedestrians, e.g. people waiting to cross at the other side of the street?\\ \hline
[None] &
[Default] How might a child, an elderly person, or someone very different from you perceive and understand this communication system? \\ \hline
\end{tabular}
\end{table}

\subsection{Pilot Study}
The initial study included a drag-and-drop task and a ranking question to evaluate preferences for eHMI placements. However, the results closely aligned with the submitted eHMI designs, making these tasks redundant, so they were removed. The sketching task initially had a 3-minute time limit, but this constraint was deemed unnecessary. Additionally, the color palette included magenta and cyan colors, which were removed in favor of using only standard colors for eHMI. 
The background scenarios used for the sketching canvas were too broad, so they were refined to direct participants’ focus on a specific area. 
Unlike the final study, this one did not include reflective follow-up questions, only using a description of the sketch and its components. 

\subsubsection{Results}
This section describes the analysis of the sketches from the pilot study, focusing on the aspects that led to key improvements in the final study. 
Out of 29 responses, 26 were analyzed, excluding 3 due to technical issues. 
The main patterns found in the sketches were as follows:

\textit{Objectives and Scope}: While participants effectively used the components from the ranking questions, their ideas were broad and lacked focus on contextual settings. To ensure more targeted responses, we narrowed the scenarios to specific characteristics and guided participants to sketch within pre-determined areas.

\textit{Challenging Ideas}: Participants often included a wide range of elements, specially auditory signals, as reported in other studies \cite{chauhan2024transforming, schieben2019designing}. However, using multiple auditory signals could create or worsen noise in the environment \cite{mahadevan2018communicating}. 
To address this, follow-up questions were designed to challenge participants’ choices and prompt them to consider alternative solutions when known limitations or potential issues are anticipated.

\section{Results}
The data from the ranking questions in the pilot study were merged with the final study data, as these questions were presented identically. 
Participants were recruited via \href{https://www.prolific.com/}{\texttt{Prolific}} and met eligibility criteria of European residence and English proficiency.  
We present the data from both the pilot study and the final study, which involved 50 participants. 
The participants had an average age of 31.31 years (SD: 10.08), with a gender distribution of 1 non-binary, 37 male, and 38 female participants.
When asked if they had seen an AV in real life, 31 answered ``No," 23 were ``Not sure," and 22 said ``Yes." 
Regarding prior interaction with AVs, 52 said ``No," 18 were ``Not sure," and 6 responded ``Yes."  
Participants were from 14 European countries, with Poland, Portugal, the UK, and Spain accounting for over half. The median study duration was 20 minutes, and participants were compensated at \pounds $12.00$ per hour for their involvement.

\begin{figure*}[t]
  \centering
  \includegraphics[width=\linewidth]{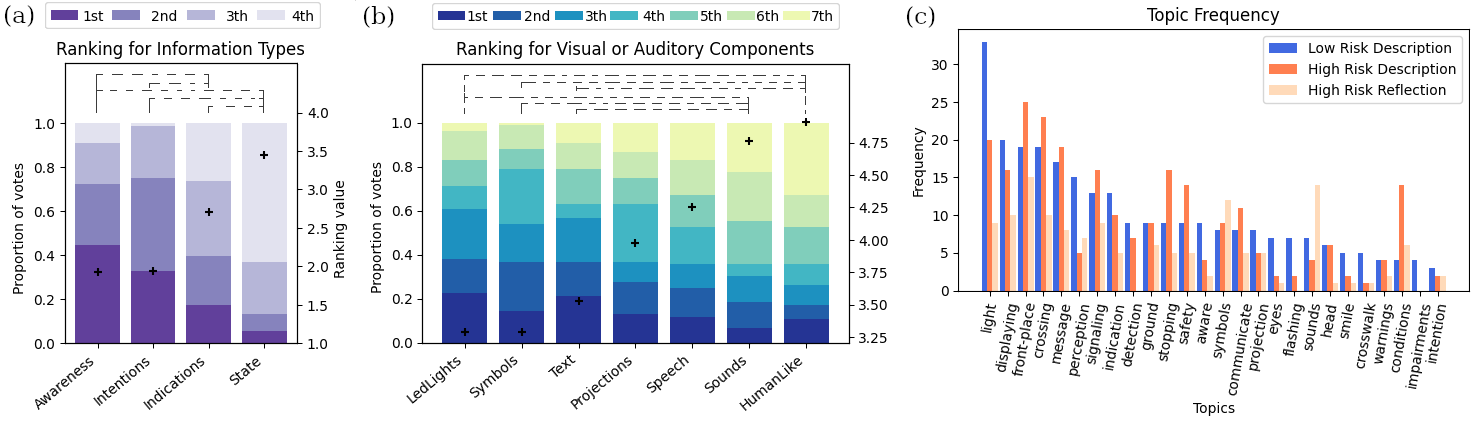}
  \caption{(a-b) Preference results for \textit{Information Type} and \textit{Visual or Auditory Components}. Plus marker (+) indicates mean ranking value. For clarity, only statistical differences at $p < 0.001$ are shown. (c) Topic Frequency shows relevant subjects participants mentioned in sketch descriptions and reflections.}
  \label{fig:rankings}
\end{figure*}

\subsection{Ranking Questions: Participants' Preferences}
All following initial statistical tests were performed with a \textit{Kruskal-Wallis rank sum test}, as the data was not normally distributed. 
Further pairwise comparisons were conducted using \textit{Wilcoxon rank sum tests with continuity correction}. 

The results for the participants' preferences on the type of \textit{information to communicate} with eHMIs indicate a similar preference for \textit{Awareness} (M: 1.92, SD: 1.00) and \textit{Intentions} (M: 1.93, SD: 0.78), followed by \textit{Indications} (M: 2.69, SD: 1.04) and \textit{State} (M: 3.45, SD: 0.85). 
The statistical test revealed a significant effect of information type on rankings, $\chi^2(3) = 84.829, p < 2.2e-16$.  
Pairwise comparisons are shown in Figure \ref{fig:rankings} for statistical differences with $p < 0.001$.

Regarding preferences for \textit{visual or auditory components}, \textit{LED lights} (M: 3.28) and \textit{Symbols} (M: 3.28) were the most favored, followed by \textit{Text} (M: 3.52).  \textit{Projections} received moderate preference (M: 3.97), while \textit{Speech} (M: 4.25), \textit{Sounds} (M: 4.76) and \textit{Human-like} (M: 4.90) components were ranked the lowest.
The statistical test indicated a significant effect of component type on rankings, $\chi^2(6) = 52.969, p = 1.19e-09$. Pairwise comparisons revealed several significant differences, those with $p < 0.001$ shown in Figure \ref{fig:rankings}-(b). 


\subsection{Sketches: Participants' Designs}

\subsubsection{Low Risk - Designs}
In the low-risk scenario, the most common element proposed by participants was the use of lights, which followed conventional color interpretations. Green was used to signal safe instructions, yellow indicated a warning or slowing down, and red associated with stopping. Some participants also used blue to convey awareness. These lights were often placed where headlights are traditionally located and employed for projections onto the road. The projections were commonly designed to create a zebra crossing, guiding pedestrians safely. 
The second most frequent element was text-based messaging, which primarily confirmed pedestrian detection or provided clear instructions. 
Additionally, some participants incorporated symbols, such as smiley faces or eyes, to create a sense of safety, as well as icons like pedestrian symbols to indicate awareness. A few participants suggested using sound signals in combination with visual cues. 
Overall, there was a strong emphasis on multi-modal communication, with many participants stressing that text should be simple, universally understood, and accompanied by symbols that are instantly recognizable, particularly for those who may not be able to read the language.

To corroborate this analysis, we plot the frequency of these topics\footnote{Responses were analyzed using Term Frequency - Inverse Document Frequency (TF-IDF) to identify relevant terms, which were then manually grouped into topics by the first author. Employed the Scikit-learn API \cite{sklearn_api}} in Figure \ref{fig:rankings}-(c). This shows that concepts like ``light" ---in particular---, ``displaying", and ``message" are frequent topics used in the original low-risk concepts. 

\subsubsection{Low Risk - Reflection}
Since visual elements played a significant role in most low-risk designs, 35 participants were asked to reflect on the effectiveness of these displays in different conditions (see Table \ref{tab:followupQuestions}-[a]). Many suggested increasing brightness to ensure visibility in foggy and rainy conditions, with some proposing adaptive brightness control. Others recommended flashing lights to enhance visibility under varying lighting conditions. Participants who incorporated multiple communication elements in their designs tended to advocate for redundancy through multiple modalities but cautioned against excessive complexity. Some participants specifically focused on the addition or enhancement of sound signals to complement visual cues.

Regarding text elements (Table \ref{tab:followupQuestions}-[c]), five participants reflected on improvements like large, high-contrast text and symbols to increase readability. Two participants were asked to reflect on audio cues (Table \ref{tab:followupQuestions}-[b]), suggesting the use of standardized sounds and adjusting audio for noisy environments (reacting to ambient noise levels). Additionally, two participants received a follow-up question on human-like features (Table \ref{tab:followupQuestions}-[d]), suggesting simple gestures like a thumbs up or smiles to improve clarity. Finally, when considering the perspective of different road users, six participants reiterated the importance of universal symbols, emphasizing the use of familiar vehicle signals to create intuitive communication.  

\subsubsection{High Risk - Designs}
In the high-risk scenario, participants applied a similar multi-modal communication approach to convey awareness and caution. Text messages were commonly placed on the windshield, while pedestrian icons, traffic-like signals, and human figures were used to indicate detection. Several designs also incorporated projected lights onto the road to create a crosswalk, reinforcing pedestrian safety. A few participants suggested dynamic signal adjustments based on situational volume and environmental conditions, such as increasing brightness in crowded areas to enhance visibility.
This analysis is reflected in Figure \ref{fig:rankings}-(c) where the topics of ``crossing", ``front-placement", ``messages", ``stopping" and ``conditions" see a notable increase. 
 
\subsubsection{High Risk - Reflection}
Finally, participants were asked to reflect on how they would modify their designs to communicate with multiple pedestrians (Table \ref{tab:followupQuestions}-[f]). Many suggested directing visual elements toward different pedestrian locations and adjusting their designs accordingly. For instance, some recommended adding arrows, bars, or pedestrian icons on display screens to show that the autonomous vehicle (AV) had detected pedestrians in various positions. Several participants also expressed interest in incorporating sound effects or traffic-light-like systems, though they did not specify how these sounds would be directed. 


\section{Discussion}
Involving everyday users in the design of eHMIs is crucial for ensuring the successful adoption of autonomous vehicles. Unlike technical experts, who may focus on cutting-edge innovations, everyday users provide invaluable insights into how communication systems should align with intuitive, familiar mental models, as depicted in the most common proposed concepts shown in Figure \ref{fig:teaser}. 

Regarding the insights gained from user-driven designs (\textbf{RQ1}), our findings reveal that participants consistently favored multimodal communication that builds on \textit{familiar traffic conventions}, such as enhanced headlights and universally recognized symbols. Rather than prioritizing novel technological solutions, participants emphasized the need for \textit{standardized signaling elements} and \textit{adaptive systems} that function effectively in diverse environments. This underscores the importance of designing for users' real-world expectations rather than solely for technical advancements. 

This user-centered preference became even more evident when considering how communication mechanisms should be modified to account for different levels of perceived risks (\textbf{RQ2}). Participants strongly preferred communication methods requiring minimal learning, often favoring a combination of lights, simple text messages, and recognizable symbols that transcend language barriers. 
By centering the needs of everyday users, eHMI design can achieve a higher level of accessibility and effectiveness.

Furthermore, the larger participant pool enabled us to support our findings with both quantitative and qualitative evidence. 
Statistically significant differences emerged in specific preferences for eHMI design elements, while participant sketches and responses provided concrete visualizations of how everyday users interpret these components. 
For instance, while previous research has outlined general guidelines for eHMI design \cite{schieben2019designing}, our study offers specific insights into how pedestrians expect these elements to function. 
Participants emphasized that text should be concise, icons should closely resemble familiar symbols (such as standard pedestrian graphics), and all elements, including LED lights and projections, should be adaptable and directed toward pedestrians.

Moreover, our web-based participatory approach proved highly effective in capturing diverse user perspectives. A key advantage was its ability to gradually familiarize participants with eHMI concepts, ensuring their input was both informed and authentic. Additionally, incorporating focused scenarios, designated drawing areas, and follow-up questions---as an improvement over the pilot test---enhanced participant responses.  
Studies show that adding structure to design tasks improves feedback quality, but overly rigid goals can limit creativity and depth \cite{morschheuser2017gamified}. Further research, possibly through collaborative methods \cite{wallace2020sketchy, park2013crowd}, is needed to balance guidance and creative freedom in participatory eHMI design.

A common concern with crowdsourced data is the possibility of minimal-effort responses, potentially leading to less innovative ideas. However, aside from a few exceptions, we found no strong evidence of this pattern in our results. Participants followed the provided guidelines carefully and were detailed in their responses (using an average of 22 words to express their ideas) ---demonstrating a meaningful level of engagement. Furthermore, all participants were willing to participate again, with many appreciating the opportunity to contribute, consistent with previous creative crowdwork \cite{oppenlaender2020creativity}. These findings suggest that well-structured participatory approaches can generate valuable user-driven insights.

It is important to emphasize that our approach is not intended to replace traditional participatory methods. Instead, we see it as a \textit{complementary strategy} that offers a parallel way to gather deeper insights and rich descriptive data from larger participant pools. This method is especially valuable for rapidly evolving technologies like autonomous vehicles, allowing everyday users to contribute to the development process. This broader reach can help \textit{clarify, validate, and expand upon} the findings obtained through conventional methods such as interviews, focus groups, or other participatory techniques. 

Our findings emphasize that everyday users must be the focal point of eHMI design for autonomous vehicles. Their intuitive understanding of traffic communication, preference for standardized and adaptable elements, and need for immediate clarity make them essential contributors to the development of effective autonomous vehicle signaling systems. By embracing a user-centered approach, we can ensure that eHMI solutions are not only technologically advanced but also universally accessible and intuitive.
 
\section{Limitations}
While our crowdsourced approach provides valuable insights into user perspectives on eHMI design, we acknowledge several limitations that should be considered when interpreting our findings and planning future research in this area. 
One key challenge---especially when using perceived risk as a factor---is that participants may struggle to fully imagine these scenarios. More interactive methods, such as video stimuli or virtual reality, could help provide better context. 
Additionally, the thematic analysis of the designs was carried out solely by the first author, which may have introduced bias and also limited the number of participants we could include in the study. As the number of participants increases, manually analyzing sketches becomes more time-consuming and less practical. To address this, future work will aim to automate the analysis process. 
Finally, while our method is designed to guide participants' creative input by building on previous research, integrating expert feedback into the process is necessary to ensure practicality and effectiveness. For instance, while \textit{redundancy} in communication is essential for clarity, experts caution against overloading users with excessive signals \cite{dey2020taming}. To address this, we plan to enhance our follow-up questions using Language Models that can analyze participant responses and generate expert-verified feedback, ensuring a balanced approach between clarity and usability.

\section{Conclusions}
Our study highlights the essential role of everyday users in shaping effective eHMI designs for autonomous vehicles. Their reliance on familiar traffic conventions, preference for standardized and adaptive elements, and need for immediate clarity underscore the importance of prioritizing user-centered design over purely technical advancements. Our participatory approach successfully captured diverse perspectives, demonstrating that well-structured crowdsourcing methods can yield meaningful and engaged contributions. 
Ultimately, designing with everyday users in mind will be key to ensuring seamless and universally understood communication between autonomous vehicles and pedestrians.

\addtolength{\textheight}{-6cm}   


\section*{ACKNOWLEDGMENT}
We would like to express our gratitude to Katie Winkle for providing valuable suggestions, and Alessio Galatolo and Hong Wang for reviewing a draft version. 
This research was supported by the Horizon Europe EIC project
\href{https://symaware.eu}{SymAware} under the Grant Agreement No. 101070802.

\bibliographystyle{IEEEtran}
\bibliography{IEEEabrv,references}

\end{document}